\documentclass[10pt,twocolumn,showpacs,amsmath,amssymb,aps,nofootinbib,floatfix,preprintnumbers,notitlepage,superscriptaddress,prx]{revtex4-2}
\usepackage{graphicx,tabularx}
\usepackage{amsmath,amssymb,amsthm,amsfonts}
\usepackage{color}
\usepackage{multirow}
\usepackage{booktabs}
\usepackage{url}
\usepackage{comment}
\usepackage{enumitem}
\usepackage{braket}
\usepackage{slashed}
\usepackage[normalem]{ulem}
\usepackage{scalerel}
\usepackage{wrapfig}
\usepackage{cancel}
\usepackage{xcolor}
\usepackage[T1]{fontenc}
\usepackage[utf8]{inputenc}
\usepackage{orcidlink}
\newcommand{\be}{\begin{eqnarray}}

\newcommand{\ee}{\end{eqnarray}}

 \theoremstyle{plain}
\newtheorem{lem}{Lemma}
\newtheorem*{lem*}{Lemma}

\newtheorem{prop}{Proposition}
\newtheorem*{prop*}{Proposition}

\newtheorem*{cor*}{Corollary}

\theoremstyle{definition}

\def\be{\begin{equation}}
\def\ee{\end{equation}}

\usepackage{hyperref}
\hypersetup{
colorlinks=true,
linkcolor=blue,
filecolor=magenta,
urlcolor=blue,
citecolor=blue,
pdftitle={Sharelatex Example},}

\begin{document}

\title{Gibbsian Entropy for General Wigner Distributions}

\author{Gabriel Rabelo-Soares \,\orcidlink{0000-0003-3843-4747}}
\email[Corresponding author: ]{gabriel.r.soares@dac.unicamp.br}
\affiliation{Universidade Estadual de Campinas - Instituto de Física Gleb Wataghin\\
Rua Sérgio Buarque de Holanda, 777\\
 CEP 13083-859 - Campinas - São Paulo - Brazil}
\affiliation{Institute of Nuclear Physics, Polish Academy of Sciences, ul. Radzikowskiego 152, 31-342 Krakow, Poland}

\author{Yevgeniy Kovchegov \,\orcidlink{0000-0001-6332-9741}}
\email[Corresponding author: ]{yevgeniy@unicamp.br}
\affiliation{Universidade Estadual de Campinas - Instituto de Matemática, Estatística e Computação Científica
\\
Rua Sérgio Buarque de Holanda, 651 \\
CEP 13083-859 - Campinas - São Paulo - Brazil}

\author{Giorgio Torrieri \,\orcidlink{0000-0002-0611-766X}}
\email[Corresponding author: ]{torrieri@ifi.unicamp.br}
\affiliation{Universidade Estadual de Campinas - Instituto de Física Gleb Wataghin\\
Rua Sérgio Buarque de Holanda, 777\\
 CEP 13083-859 - Campinas - São Paulo - Brazil}
\begin{abstract}
We propose a definition of non-equilibrium Gibbsian entropy applicable to general quantum systems, computable from Wigner functions, for both Gaussian and non-Gaussian states.  We show that while the Wigner function is a quasi-probability distribution, fundamental results linking observables in repeated observations to the function, such as the law of large numbers, the central limit theorem and, crucially, the relation between the entropy and the cumulant generating function continue to apply. This ideas are the core of the large deviation theory, which allow to formulate a positive-definite definition of entropy based on our ignorance of the underlying state structure after repeated measurements of cumulants. This new definition is a parallel to the Gibbs entropy formulation, but it can also be used in out-of-equilibrium regimes. Some important properties of this entropy are discussed. In addition we derive some information theory quantities such as relative entropy, conditional entropy, mutual information and Rényi entropy. We then compare the results with Wehrl entropy for the 1D harmonic oscillator, for pure and mixed states. We conclude the the work discussing potential applications of this new formalism, ranging from quantum correlations beyond entanglement to thermalization in closed quantum systems. 

\end{abstract}
\maketitle

\tableofcontents

\section{Introduction}
The concept of entropy was originally introduced in classical thermodynamics, being a state function that characterizes the irreversibility of physical processes. With the development of statistical mechanics, Ludwig Boltzmann established a microscopic interpretation of this quantity by relating the thermodynamic entropy to the number of microscopic configurations compatible with a given macroscopic state. This relation was expressed by the celebrated Boltzmann formula, ($S_B=k_B\ln W$), with $W$ the number of accessible microstates. The usefulness of entropy as a physical quantity is based around the idea that phase-space configurations away from equilibrium are exceedingly rare.  Within rigorous probability theory rare events are quantified via large deviation principle and it can be shown that statistical mechanics is directly related to this principle \cite{touchette}. 

Subsequently, the statistical description of entropy was generalized by J. Gibbs through a formulation based on the probability distribution over the system's phase-space. In the classical framework, the Gibbs entropy is defined as $S_G=-k_B\int d\Gamma,\rho(\Gamma,t)\ln\rho(\Gamma,t)$, where ($\Gamma$) denotes a point in phase-space and ($\rho(\Gamma,t)$) the normalized phase-space probability density. While Boltzmann's formulation relates entropy to the multiplicity of microstates compatible with a macrostate, Gibbs' formulation explicitly accounts for their probability distribution, recovering the Boltzmann expression when the accessible microstates are equiprobable (provided a consistent phase-space measure is adopted). The statistical formulation also establishes a natural connection between entropy and information theory \cite{shannon1948mathematical}, in which entropy can quantifies the uncertainty associated with a probability distribution.

These entropic definitions are not directly applied to the quantum mechanics framework as will be discussed next. A general quantum system can be described by a density matrix $\rho$ operator, which has features like being positive semi-definite, Hermitian, and an unity trace. Based on this operator, von Neumann introduced a generalization of the Boltzmann entropy \cite{VonNeumann1955}, defined as: $S_{\text{vN}} = - \text{Tr} \rho \ln \rho$. In the case where the density matrix has positive eigenvalues, i.e., $\lambda_i \ge 0$ and $\sum_i \lambda_i = 1$, the von Neumann entropy can be expressed in terms of the Shannon entropy $S_{\text{vN}} = - \sum_i \lambda_i \ln \lambda_i$.
The Shannon formulation lead to an informational interpretation, for $S_{\text{vN}} \ge 0$, with the equality holding when one has access to all information about the system, or, equivalently, when the system in question is in a pure state. Consequently, the von Neumann entropy could be used as a measure of the degree of mixedness of the quantum system.

In the same way one may wonder about the analogue of the classical phase-space entropy in quantum mechanics. In the phase-space formulation of quantum mechanics, quantum states and quantum operators are described by continuous functions of the canonical variables of momentum and position, $p$ and $x$, respectively. The Wigner-Weyl transform \cite{WignerWeyl2008} is the map of any linear operator $O$ into a distribution $O(x,p)$. From this transformation, one can then define the Wigner distribution $W(x,p)$ from the density matrix of the system. The phase-space representation of quantum mechanics has emerged as a powerful approach to study and intuitively describe quantum phenomena, being useful to distinguish classical from quantum structure (also quantum and classical dynamics), non-Gaussianity, quantum optics among others \cite{Moyal_1949, HILLERY84, Simon87, PhysRevA1991, PhysRevA95, Kenfack_2004, Illuminati2006, Paris2008, Leo2009, Mari2012, PRA2018, nongaussian}. 

The Wigner distribution have many of the properties that one expects for a phase-space probability distribution, i.e., it is real, normalized to unity, gives the right marginal distributions of momentum and position when integrated over position and momentum respectively \cite{Wignermarginals1980}. Wigner distributions can also be used to compute averages of any observable ($\braket{O} = \int dx dp W O$). Although these properties, Wigner distributions cannot be interpreted as a genuine probability distribution as it can assume negative values \cite{Kenfack_2004, PRA2018}, regarding the correlations between the position and momentum constrained by the uncertainty principle which reflects the non-classicality of the quantum states. 

Therefore the quantum phase-space entropy is an ill-defined quantity for the non Gaussian states. The Shannon entropy of the Wigner distribution can only be defined for the Gaussian Wigner states as studied previously in \cite{Manfredi_2000, PRA2021}. Hudson's theorem \cite{HUDSON1974249} sets that Gaussian pure states are the only one with positive Wigner distribution, one the side of mixed states, however, there are some efforts in the description of mixed Gaussian positive states and even a extension of Hudson's theorem \cite{mixed_wigner_1995, mixed_Hudson_2009}. As such, this entropy captures the uncertainty of the quantum phase-space in a information theory point of view. It is worth noting, nevertheless, that because it concerns a quantum distribution, it is subject to the uncertainty principle of the quadrature variables, consequently, this entropy has a lower bound, known as the Wigner entropy conjecture \cite{Manfredi_2000, Hertz_2017, PRA2021, quantentropy, PRA2024, Van_Herstraeten_2025}. Even for the select class of Gaussian Wigner states, the Wigner entropy have many applications, for instance in the context of non equilibrium thermodynamics in order to measure the entropy production in quantum systems \cite{Santos_2017, PhysRevLett.121.160604}, harmonic oscillator models \cite{PhysRevA71, PRA2023}. It was shown in \cite{PhysRevLett2012} that the Wigner entropy can be written in terms of the Rényi entropy of order 2 relating it to the purity of the system, which will be revisited later on this work. 

For the said non Gaussian states, a recent work  \cite{quantentropy} extended the Wigner entropy to the case of negative Wigner values through the analytic continuation of Shannon’s differential entropy. However, for the means of this work, we will define and Wehrl entropy as the standard procedure to compute the entropy of non-Gaussian states. For the negative Wigner distribution, on can define the Husimi distribution, i.e., a gaussian smearing of the Wigner distribution that constitute a positive definite distribution in which the Wehrl entropy \cite{wehrl} is defined on. As the Husimi distribution depends on a smearing factor, the entropy is not uniquely defined. Furthermore, the Gaussian smoothing of the Wigner distribution spread the intrinsic quantum correlations over the phase-space, which is responsible for the semiclassical nature of the Wehrl entropy.

Although this arguments Wehrl entropy has been shown to be useful in some contexts, powered to the fact that Hisimi distribution can be easily experimentally accessible \cite{schleich2001quantum} (Wigner distributions can also be accessed experimentally even in the presence of negativity states \cite{Kurtsiefer1997, Haroche2000, Clerk18, Winkelmann_2022}). For Wehrl entropy, it was demonstrated that it can be used as a Witness of entanglement, where it sets the lower bound of the entanglement in the system \cite{PRA2004, FloerchingerHaasMueller2021}. This entropic formulation is also useful in many scenarios as the density matrix is not always known for a given system, making the von Neumann entropy hard to be defined or even computed. Finally, in a recent work \cite{Tang_2025} the authors have explored the Wehrl entropy and its connection to complexity in quantum systems.

Many progress was done in the direction of the phase-space entropy, both for Wigner distribution entropy for the Gaussian states, and Wehrl entropy for smooth phase-space distributions. Nevertheless, the non-Gaussianity is not proper explored and understood in terms of the entropy measures. In quantum mechanics framework, the non-Gaussianity and the negativity of the Wigner function have been explored extensively \cite{Kenfack_2004, Paris2008, Leo2009, PRA2018, nongaussian}. This work aims to give a first view on this matter by proposing a entropy definition that holds for both Gaussian and Non-Gaussian states without the necessity of defining a smooth phase-space distribution.

To address this goal, we employ the large deviations theory \cite{Dembo10}, a topic of the probability theory, which is dedicated to study and understand the asymptotic behavior of the tails of a probability distribution, or equivalently, the exponential decay of the probability of large fluctuations in a given random system \cite{Cramer1938,DonskerVaradhan1975I, DonskerVaradhan1975II, DonskerVaradhan1976III, DonskerVaradhan1983IV, FreidlinWentzell1984, Ellis1999}. Although its formal development is recent, it is argued in the literature that the first result in large deviation theory appeared within the formulation of statistical mechanics \cite{Ellis1999}. While still relatively under-explored in physics, this theory has gained attention within the community. It has proven to be a fundamental mathematical framework for statistical mechanics, appearing for instance, whenever entropy or free-energy functions are calculated \cite{touchette}.

This work draws upon the connection between large deviation theory and statistical mechanics to formulate the entropy of a Wigner distribution that we named by large deviation principle entropy $S_{\text{LDP}}$. Since the large deviation theory provides a collection of methods for calculating cumulant-generating functions (analogous to free energies) and entropy rate functions (analogous to entropy), this work focuses on defining these quantities for the Wigner distribution. An added advantage is that, within this formalism, the Wigner distributions do not need to be strictly Gaussian, thereby demonstrating the potential application and utility of these results in the non-Gaussian sector not yet explored in the context of quantum phase-space entropy. Finally, we explore this definition by highlighting some of its fundamental properties, as well as we take as an example the 1D harmonic oscillator to compare the different definitions of phase-space entropy. We also derive some information theory quantities such as mutual information, relative entropy, conditional entropy and also Rényi entropy for $S_{\text{LDP}}$. Finally, we explored the entropic uncertainty relation for the large deviation principle entropy making the comparison with the uncertainty relations already defined in the literature. 

Accordingly, the remainder of this work is organized as follows: in Section \ref{sec:Wigner_Phase_Space_Distribution_Large_Deviation_Construction} we connect the large deviation theory with Wigner distribution, defining then its log-momentum generating function and rate function. In Section \ref{sec:Gibbsian_Entropy_Wigner_Rate_Function} we introduce the proposal for the entropy formula, consequently some of it properties, we also study the central limit theorem as a limit of the large deviation principle. In Section \ref{sec:Harmonic_Oscillator_Analytical_Numerical_Results} we explore the 1D harmonic oscillator both the pure and mixed states in order the make a comparison of the entropic definitions. In Section \ref{sec:Information_Theory_Induced_Wigner_Rate_Function} we explore some information theory quantities induced by the large deviation entropy as the relative entropy, conditional entropy, mutual information and Rényi entropy. In Section \ref{sec:Entropic_Uncertainty_Quantum_Correlations} we study the entropy uncertainty relation derived from our entropy definition and its relations to the uncertainty relations already established in the literature, we finish the section by relating relating the mutual information as a witnesses of entanglement for pure states. Finally, in Section \ref{sec:Discussion_Outlook} we discuss the main results of the present work and explored the potential applications of this entropy across the different scenarios, as thermalization in closed systems, quantum correlation beyond entanglement and non-Gaussianity. At last, in Section \ref{sec:conclusions} we present the final conclusions. Some Appendix are also included in order to leave the main text with the central results only. 

\section{Wigner Phase-Space Distribution and the Large-Deviation Theory}

In this section, we will restrict ourselves to the formulation and connection of the Wigner distribution within the framework of large deviation theory. Specifically, we will show how to formulate the moment-generating and cumulant-generating functions for $W(x,p)$. Consequently, the large deviation principle can be stated, ultimately enabling the calculation of the rate function associated with the Wigner function. These steps are necessary for the subsequent formulation of the entropy. For the purposes of this work we are using natural canonical coordinates with $[x,p]=i$, natural logarithms and $\hbar =1$ unless states otherwise. 

\label{sec:Wigner_Phase_Space_Distribution_Large_Deviation_Construction}
\subsection{Wigner distribution and its Moments}

For a pure state, the Wigner function for a given wave function $\psi(x)$ ($\psi(p)$) in coordinate (momentum) space is given by its Weyl transform \cite{WignerWeyl2008}
\begin{equation}
    W_{\psi} (x,p) = \frac{1}{2 \pi} \int_{\mathbb{R}} \,dy\, e^{ipy} \, \psi^{*}(x+ y/2) \psi(x- y/2).
    \label{eq:wignerdef}
\end{equation}
For any $m, n \in \mathbb{N}$, the corresponding $(n,m)$-moment of $W(x,p)$,
\begin{equation}
    \label{eqn:McCoyi}
 \iint\limits_{\mathbb{R^2}} x^n p^m \, W(x,p) \,dx\, dp
={1 \over 2^m}\sum\limits_{j=0}^m {m \choose j} \langle {\sf p}^j {\sf x}^n {\sf p}^{m-j} \rangle,
\end{equation}
known as the \textit{McCoy's formula}, the detailed derivation of this formula can be found in Appendix \ref{app_moments_charc_function}. Yet on Eq. (\ref{eqn:McCoyi}), let $\widehat{\widehat{W}}(t,k)=\mathcal{F}_{x,p}[W](t,k)$, denote the multi-dimensional Fourier transform of $W(x,p)$. Since $W(x,p)$ is itself an inverse Fourier transform in the momentum space, we have that Eq. (\ref{eqn:McCoyi}) implies,
\begin{equation}
\label{eqn:WhhMc}
\widehat{\widehat{W}}(t,k)= \iint\limits_{\mathbb{R}^2} \! \,dx\, dp \, e^{-itx-ikp} \, W(x,p) 
=\langle e^{-ik{\sf p}/2} e^{-it{\sf x}} e^{-ik{\sf p}/2} \rangle ,
\end{equation}
$\forall t,k \in \mathbb{R}$. Thus the Wigner representation supplies statistical moments without furnishing an ordinary joint probability for simultaneous sharp position and momentum measurements.
   
\subsection{Moment Generating Function}

The previous formulation are used then to define the (log)-momentum generating function. From Eq. \eqref{eqn:McCoyi} we can define {\it moment generating function} of $W(x,p)$:
\be\label{eqn:mgf}
M_W(t,k)= \iint\limits_{\mathbb{R}^2} \! e^{tx+kp} \, W(x,p) \,dx\, dp
=\langle e^{k{\sf p}/2} e^{t{\sf x}} e^{k{\sf p}/2} \rangle,
\ee
$\forall t,k \in \mathbb{R}$. Observe that, by symmetry, $M_W(t,k)>0$ for all $t,k \in \mathbb{R}$. Hence, for the log-momentum generating function we introduce the following lemma.\begin{lem}\label{lem:convex}
The {\it log-moment generating function} of $W(x,p)$,
$$\log M_W(t,k)=\log\langle e^{k{\sf p}/2} e^{t{\sf x}} e^{k{\sf p}/2} \rangle,$$  
is convex.
\end{lem}
\medskip
\noindent
The function $\log M_W(t,k)$ is also called the {\it cumulant generating function} (see Appendix \ref{app_cumulant_convexity} for a  proof of this Lemma).

\subsection{Large Deviation Principle}

Being defined the log-momentum generating function we can introduce the large deviation principle in general terms. Consider a sequence of independent identically distributed (i.i.d.) random variables $\xi_1,\xi_2,\hdots$ in $\mathbb{R}^d$ and let
$S_n= \sum\limits_{j=1}^n \xi_j$.
For a set $A \subseteq  \mathbb{R}^d$ such that $O \subseteq A \subseteq \bar{O}$, where $O \subseteq  \mathbb{R}^d$ is an open set
and $\bar{O}$ is its closure, the {\it large deviations principle} (LDP) states that 
\be\label{eqn:LDPc}
-\lim\limits_{n \to \infty}{1 \over n}\log P\left({S_n \over n} \in A\right)=\inf\limits_{a \in A}\mathcal{I}(a).
\ee
By Carm\'{e}r's Theorem, further generalized in the Gartner-Ellis Theorem, the large deviations {\it rate function} $\mathcal{I}(a)$
is the Legendre-Fenchel transform of the log-moment generating function $\log M(a)$ of $\xi_i$,
\begin{equation}
\label{eq:LDTrate}
    \mathcal{I}(a)=\sup_{{\bf x} \in \mathbb{R}^d}\big\{\langle a, {\bf x}  \rangle - \log M(a)  \big\},
\end{equation}
where $\langle \cdot, \cdot \rangle$ denotes the inner product in $\mathbb{R}^d$. The above equation appears many times in the context of statistical mechanics and thermodynamics, and is where the success of the large deviation theory (LDT) as a fundamental mathematical formulation of these theories lives in. In other words, LDT gives an formal explanation of why entropy and free energy functions are connected via the Legendre transform (for more details regarding the connection between large deviation theory and statistical mechanics, see \cite{touchette}). 

\subsection{Rate Function Associated with the Wigner Distribution}
In that matter, we can proper define now the large deviation theory results in Eqs. (\ref{eqn:LDPc}) and (\ref{eq:LDTrate}) in term of the Wigner distribution. Lemma \ref{lem:convex} ensures the validity of defining the {\it large deviations entropy rate} of $W(x,p)$ via the Legendre transform 
\begin{equation}
    \begin{aligned}
    \label{eqn:entropy}
I_W(x,p)=\sup\limits_{t,k \in \mathbb{R}}\big(xt+pk -\log M_W(t,k)\big).
\end{aligned}
\end{equation}
The large deviation principle, as well as the Lemma \ref{lem:convex}, ensures and guarantee that Wigner rate function $I_W(x,p)$ is a concave function. This results are directly extended to Husimi distributions, which is described in more details in the Appendix \ref{app_Husimi_LDP}. 

\section{A Gibbsian Entropy from the Wigner Rate Function}
\label{sec:Gibbsian_Entropy_Wigner_Rate_Function}
\subsection{Large Deviation Entropy}
While the large deviation principle Eq.~\eqref{eqn:LDPc} is often loosely interpreted as
$$P\left({S_n \over n} \in A\right) \approx e^{-n\, \mathcal{I}(A)}, \quad  \text{ where } ~~\mathcal{I}(A):=\inf\limits_{a \in A}\mathcal{I}(a),$$
in \cite{BorovkovMogulSkii1996, Borovkov2013}
and related works, the authors refined Eq.~\eqref{eqn:LDPc} to yield
\be\label{eqn:Borovkov}
P\left({S_n \over n} \in A\right) \approx {n^{d/2} \over (2\pi)^{d/2}}\int\limits_A e^{-n\, \mathcal{I}(a)} {da \over \sigma(a)},
\ee
for all bounded measurable $A \subseteq \mathbb{R}^d$, where 
$$\sigma^2(a)=\det E\big[(\xi^{(a)}-a)^T (\xi^{(a)}-a)\big],$$
with $\xi^{(a)}$ distributed according to the following Carm\'{e}r transform of the original distribution of $\xi_i$'s,
$$P(\xi^{(a)} \in B)={E[e^{\langle \xi_i, \mathcal{I}'(a) \rangle} ;\,  \xi_i \in B] \over M\big(\mathcal{I}'(a)\big)}.
$$
Both Boltzmann entropy and the large deviations rate function measure the exponential decay rate for the probability of rare events. As an ad hoc approach to converting the rate function $\mathcal{I}(a)$ into a Shannon entropy, we consider the probability density function
\begin{equation}
\label{eq:probdensityLDP}
    {1 \over \mathcal{Z}}e^{-\mathcal{I}(x)}, \quad \text{ where } ~~\mathcal{Z}:=\int\limits_{\mathbb{R}^d} e^{-\mathcal{I}(a)} \, da,
\end{equation}
which yields the Shannon entropy,
\be\label{eqn:SLDP}
S_{\text{LDP}}={1 \over \mathcal{Z}} \int\limits_{\mathbb{R}^d} \mathcal{I}(x) e^{-\mathcal{I}(x)} \, dx +\log \mathcal{Z}.
\ee
This is the expression for the entropy of the Wigner distribution, based on the theory of large deviations. Observe that under a linear symplectic transformation $(x',p')=T(x,p)$ the large deviation entropy changes from $I_W(x,p)$ to $I_W\big(T^{-1}(x,p)\big)$. Since the determinant of the matrix for $(x',p')=T(x,p)$ is equal to one, the integrals $\mathcal{Z}=\int e^{-\mathcal{I}(x,p)} \, dx \,dp$ and $\int \mathcal{I}(x,p) e^{-\mathcal{I}(x,p)} \, dx \,dp$ are preserved under $T$. Hence, the large deviation entropy $S_{\text{LDP}}$ is invariant under symplectic transformations of the phase-space.

\subsection{The Central Limit Theorem}

In summary, the large deviation theory is concerned with the probability of rare events occurring in the tails of distributions, where the observed deviation from the mean is on the order of the sample size. Conversely, the central limit theorem (CLT) addresses the opposite scenario, where fluctuations are small and concentrated near the sample mean. In this sense, the central limit theorem can be derived from the theory of large deviations by performing a Taylor expansion of the rate function around a minimum point (i.e., the mean). Consequently, the large deviation principle (Eq.~\ref{eqn:LDPc}) yields the Gaussian kernel. Four our framework we can also study this limit (further details can be found in Appendix \ref{app_central_limit_theorem}). Therefore, being the Gaussian limit defined, it is possible to compute its entropy using the Shannon entropy relation, in which for a Gaussian distribution depends solely on the covariance matrix and can be expressed in a general form as follows:
\begin{equation}
    S_{\text{CLT}} = \frac{d}{2}(1 + \log 2 \pi  ) + \frac{1}{2} \log \text{det} \Sigma,
    \label{eq:CLT}
\end{equation}
where d represents the number of dimensions of the system and $\Sigma$ is the covariance matrix given by Eq.~(\ref{eqn:Sigma}). In the work we then define $S_{\text{CLT}}$ as the CLT entropy derived from the limit of the large deviation principle. 

The results present so far composes the theoretical construction of the large deviation entropy $S_{\text{LDP}}$ and central limit theorem entropy $S_{\text{CLT}}$. In the next section, we will discuss an example in which we compare our formulation of Wigner entropy with Wehrl entropy for the one-dimensional harmonic oscillator, covering both pure and mixed states.

\section{1D Harmonic Oscillator: Pure and Mixed states}
\label{sec:Harmonic_Oscillator_Analytical_Numerical_Results}
   \subsection{Wigner and Husimi Distributions}
   For the 1D harmonic oscillator, the Wigner and Husimi distribution are defined as, 
 \begin{equation}
    \begin{aligned}
          W_n(x,p) &= \frac{(-1)^{n}}{\pi \hbar} e^{- \frac{2}{\hbar \omega} H(x, p)} \text{L}_n \left( \frac{4}{\hbar \omega} H(x, p) \right) \\
          Q_n(x,p) &= \frac{1}{n!} e^{-\frac{H(x, p)}{\hbar\omega}}\left( \frac{H(x, p)}{\hbar \omega} \right)^{n}.
    \end{aligned}
    \label{eqn:wigner_husimi_HO}
\end{equation}
Here $L_\text{n}$ is the Laguerre polynomial and $H(x,p) =  \frac{p^2}{2m} + m \omega^2x^2$ the 1D Hamiltonian. Unlike the Wigner distribution, the Husimi distribution is nonnegative and so has a well defined entropy which is calculated from the Wehrl entropy \cite{wehrl},
\begin{equation}
    S_{W}^{(n)} = \ln \pi + n - n (\text{H}_n - \gamma) + \ln n! + 1,
    \label{eqn:Wehrlentropy}
\end{equation}
with $\text{H}_n$ is the harmonic number $\text{H}_n = \sum^{n}_{k=1}  1/k$ and $\gamma$ the Euler-Mascheroni constant. With the above expression one can easily obtain the entropies for the ground and excited states. In the sequence, we will derive the large deviation entropy for the Wigner distribution (Eq. (\ref{eqn:SLDP})) for the harmonic oscillator Wigner distribution in Eq. (\ref{eqn:wigner_husimi_HO}) to then be able to compare the results with the Wehrl entropy in Eq. (\ref{eqn:Wehrlentropy}).

Hence, to construct the large deviation entropy of the Wigner distribution we have to define the momentum generating function as well as the Legendre transform to obtain the rate function. The Legendre can be done analytically by hand in some specific cases like the ground state of the harmonic oscillator, otherwise it has to be done numerically. The derivation in the sequence are performed considering all the parameters like the mass $m$, the frequency $\omega$ and also the Plank's and Boltzmann's constants, but the plots and analysis are done considering $k_B = \hbar = m = \omega = 1$ for simplicity.

\subsection{Ground State and Excited States}
From Eq. (\ref{eqn:wigner_husimi_HO}) we can easily define the Wigner distribution for the ground state and then use Eq. (\ref{eqn:mgf}) to calculate the momentum generating function and finally, from Eq.~(\ref{eqn:entropy}), the rate function can be obtained 
\begin{equation}
    \begin{aligned}
        I_W^{(0)}(x,p) = \sup_{t,k \in \mathbb{R}} \left(  xt + pk -  \frac{\hbar t^{2}}{4 m \omega} - \frac{\hbar m \omega k^{2}}{4} \right),
    \end{aligned}
    \label{eq:Legendre_trans_ground_state}
\end{equation}
the detailed derivation can be found in the Appendix \ref{app_generating_funciton_HO}. For the ground state the supremum is straightforward and therefore the rate function can be expressed as $ I_W^{(0)} = 2 H(x,p)/\hbar \omega$.
At last, the large deviation entropy $S_{\text{LDP}}$ in Eq. (\ref{eqn:SLDP}), 
\begin{equation}
    \begin{aligned}
        S^{(0)}_{\textrm{LDP}}(W) &= \frac{1}{\mathcal{Z}} \int \int d x dp \, \, I^{(0)}_W e^{- I^{(0)}_W} + \ln \mathcal{Z} = \ln \pi + 1,
    \end{aligned}
\end{equation}
where the partition function is given by a gaussian integral in coordinate and momentum space. From the Wehrl entropy in Eq. (\ref{eqn:Wehrlentropy}), we have that its ground state value coincide with the $S_{\text{LDP}}$, been $S_W^{(0)} = \ln \pi + 1$. One my notice that for the ground state the Wigner distribution is Gaussian and positive. The more interesting analysis comes from the non gaussian states that are present in the excited states.

\begin{figure}[t!]
    \centering
    \includegraphics[width=0.9\linewidth]{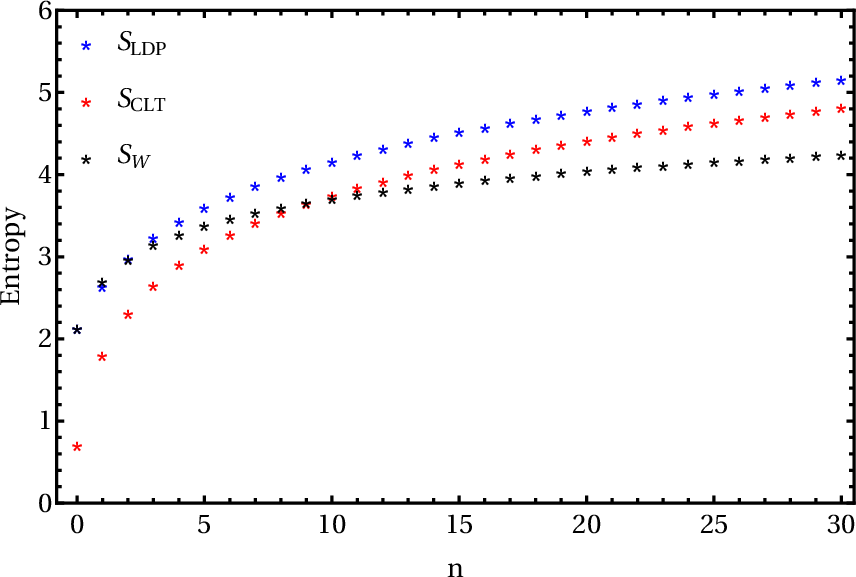}
    \includegraphics[width=0.94\linewidth]{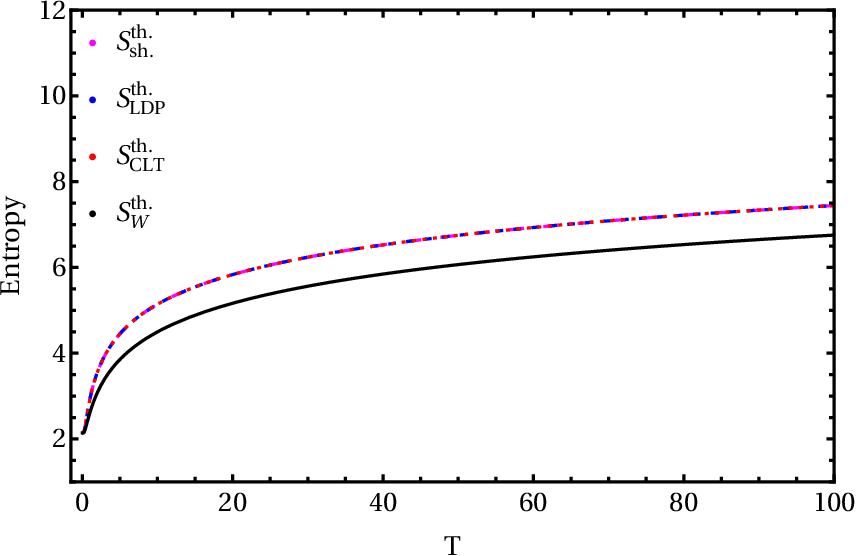}
    \caption{In this figure, we present the results for various entropies in the case of a 1D harmonic oscillator. In the top panel, we define $S_{\text{LDP}}$ and $S_{\text{CLT}}$, as well as the Wehrl entropy $S_W$, as a function of the harmonic oscillator's energy levels for pure states. In the bottom panel, we calculate the same set of entropies and, additionally, show the Shannon entropy $S_{\text{sh.}}^{\text{th}}$ for mixed states of the harmonic oscillator, as a function of the temperature.}
    \label{fig:1D_Sldp}
\end{figure}

Following the analytical expression for the momentum generating function Eq. (\ref{eq:generalMn_HO}), the cumulants generating function is straight obtained and we can then define the large deviation entropy, 
\begin{widetext}
    \begin{equation}
    I^{(n)}_W(x,p) = \sup_{t,k \in \mathbb{R}} \left( xt + p k - t^{2} \frac{\hbar}{4 m \omega} - k^{2}\frac{m \omega \hbar}{4} - \ln \left[ \text{L}_n\left(- t^{2} \frac{\hbar}{2 m \omega} - k^{2}\frac{m \omega \hbar}{2} \right) \right]\right),
\end{equation}
\end{widetext}
the exited states will be composed by logarithm terms of the n-$th$ order Laguerre polynomials. The computation of this Legendre transform will be done with help of Mathematica. The summary results are then displayed in the upper panel of Figure \ref{fig:1D_Sldp}. In the top panel, we plot the Wehrl entropy $S_W$ (Eq. \eqref{eqn:Wehrlentropy}) and the two entropies defined in this work: $S_{\text{LDP}}$ (Eq. \eqref{eqn:SLDP}) and $S_{\text{CLT}}$ (Eq. \eqref{eq:CLT}), for a one-dimensional harmonic oscillator as a function of energy levels, considering pure states only. We observe that for the ground state ($n = 0$), $S_{\text{LDP}}$ coincides with $S_W$, whereas $S_{\text{CLT}}$ is smaller. As we move toward higher excited states, the separation between $S_{\text{LDP}}$ and $S_W$ increases. This highlights our main result: even though $S_W$ is defined based on a smoothed distribution, the large deviation entropy remains larger. We interpret this as a consequence of the latter accounting for correlations and negativity of the phase-space distribution, which become more pronounced in higher excited states. As expected, $S_{\text{CLT}}$ does not coincide with $S_{\text{LDP}}$, a result consistent with the argument presented earlier, since we are employing a Gaussian limit for the excited states of the harmonic oscillator, even though these states are known to be non-Gaussian.

\subsection{Mixed Thermal State}
So far the analysis were restricted to the pure states, one other interesting analysis is regarding the entropies for the mixed states. Still on the harmonic oscillator, a straightforward example is the harmonic oscillator coupled to a thermal bath. Introducing the density operator $\rho = \sum_{n=0}^{\infty} P_{n} \ket{n}\bra{n}$ where $\ket{n}$ are the Fock states for the energy eigenstates, and $\sum_{n=0}^{\infty} P_{n} = 1$. Following from the definition of the Wigner distribution, it is straight forward do check that for a mixed state,
\begin{align}
    W_{\text{mixed}}(x,p) &= \sum_{n=0}^{\infty} P_{n} W_{n}(x,p).
    \label{wmixedI}
\end{align}
The Wigner distribution of a mixed state is a classical sum of the Wigner distribution of pure states. The weight of the density matrix describing a thermal equilibrium ensemble is given by a Boltzmann distribution
\begin{equation}
    P_{n}(E_n) = \frac{e^{- \beta E_n}}{Z},
    \label{eqn:boltzmannweightthermal}
\end{equation}
where $Z$ is the partition function. For the harmonic oscillator $E_n =  \hbar \omega \left( n + \frac{1}{2} \right)$ yielding
\begin{equation}
    P_n = e^{-\beta n}(1-e^{-\beta}). 
\end{equation}
With some algebra one can check the Wigner function for a mixed state in Eq. (\ref{wmixedI}) reduces to
\begin{equation}
    W_{\text{mixed}}(x,p) = \frac{1}{\pi\hbar} \tanh\left(\frac{\beta\hbar\omega}{2}\right) e^{-\frac{2H}{\hbar\omega} \tanh\left(\frac{\beta\hbar\omega}{2}\right)} .
    \label{eq:wignermixed}
\end{equation}
Analogously, the thermal Husimi distribution, 
\begin{equation}
    Q(x,p) = \frac{1}{\pi}\left(1-e^{-\beta} \right)e^{-\frac{H(x,p)\left(1-e^{-\beta} \right)}{\hbar\omega}}.
    \label{eqn:thermalhusimidist}
\end{equation}
For the momentum generating function the process is similar, such that the thermal version of Eq. (\ref{eq:generalMn_HO}) is
\begin{align}
    M_{W}^{th}(t,k) &= \text{Tr}(\rho_{\text{th.}} e^{\frac{tp}{2}} e^{kx} e^{\frac{tp}{2}}) \nonumber \\
       &= \sum_{n=0}^{\infty} P_{n} M_{W}^{(n)}(t,k) ,
\end{align}
with $P_n$ following from Eq. (\ref{eqn:boltzmannweightthermal}) and hence,
\begin{equation}
    M_{W}^{\text{th}}(t,k) = \exp\left[ \left(\frac{t^2\hbar}{2m\omega} + \frac{\hbar\omega m k^2}{2}\right) \frac{1}{2} \coth\left(\beta\frac{\hbar\omega}{2}\right) \right].
\end{equation}
Therefore the large deviation entropy
\begin{equation}
    I_W(x,p) = \sup_{t,k} \left( x t + p k - \frac{1}{2} \left(\frac{t^2}{2} + \frac{k^2}{2}\right) \coth\left(\frac{\beta}{2}\right) \right),
\end{equation}
considering $\hbar = m = \omega = k_B = 1$, the Legendre transform leads to
\begin{align}
    I_W(x,p) &= (p^2 + x^2) \tanh\left(\frac{\beta}{2}\right).
    \label{eqn:ThermalLDP}
\end{align}
As a final task, we can now compute the Wehrl entropy of thermal Husimi distribution Eq. (\ref{eqn:thermalhusimidist}), 
\begin{equation}
    S_W^{th} = -\ln(1-e^{-\beta}) + \ln\pi + 1,
\end{equation}
a straightforward check for $T \rightarrow 0 \Rightarrow e^{-\beta} \rightarrow 0 \Rightarrow \ln(1-e^{-\beta}) \rightarrow 0$, then $
    S_W^{th} = \ln\pi + 1 = S_W^{n=0} \nonumber$, i.e.,  the thermal Wehrl entropy reduces to the the Wehrl entropy of the ground state. For the large deviation entropy Eq. (\ref{eqn:SLDP}), for the rate function Eq. (\ref{eqn:ThermalLDP}), 
\begin{equation}
     S_{\textrm{LDP}}(\beta) =  1 + \ln \pi - \ln \tanh\frac{\beta}{2}.
\end{equation}

In the bottom panel of Figure \ref{fig:1D_Sldp}, we present the results for mixed states of the harmonic oscillator as a function of temperature. As shown in the derived equations, these thermal states are Gaussian in nature, therefore, in addition to the Wehrl, LDP, and CLT entropies, we also plot the Shannon entropy for the Wigner distribution obtained in Eq.~\eqref{eq:wignermixed}. We observe that $S^{\text{th.}}_{\text{sh.}}$, $S^{\text{th.}}_{\text{LDP}}$, and $S^{\text{th.}}_{\text{CLT.}}$ are equal to one another and greater than $S_W$, except in the case where $T = 0$. In this limit, the Wigner distribution is Gaussian and corresponds to the ground state of the harmonic oscillator.

Our main finding of this section is that, for the excited states of a harmonic oscillator, the large deviation principle (LDP) entropy is greater than the Wehrl entropy. This may seem counterintuitive, given that the construction of the Wehrl entropy, based on a Gaussian smoothing of the Wigner distribution, might suggest that it serves as an upper bound for entropy. We argue, however, that the LDP entropy is higher precisely because it accounts for the non-Gaussian nature of the harmonic oscillator excited states, where quantum correlations and negativities contribute to the observed entropy. Finally, we note that we have restricted our comparison to the Wehrl entropy. The Wigner entropy, as defined in \cite{Manfredi_2000, PRA2021}, is formulated for Gaussian states, such as the vacuum state, and, in that context, yields the same value as the LDP entropy. Next, we will also present some results, such as relative entropy, conditional entropy, mutual information, and Rényi entropies derived from the large deviation entropy. 

\section{Information Theory Induced by the Wigner Rate Function}
\label{sec:Information_Theory_Induced_Wigner_Rate_Function}

Throughout this section we will then derive some other quantities inferred from the large deviation entropy of the Wigner distribution. Among these quantities, we focus on relative entropy, conditional entropy, mutual information and Rényi entropy, which have already been previously explored for Wigner entropy (for Gaussian states), Wehrl entropy and density matrix \cite{adami97, Paris2008, FloerchingerHaasMueller2021, PRA2023}.  
\subsection{Relative Entropy}
   Considering two rate functions $I_P$ e $I_Q$, obtained, for example, from two different quantum states, the relative probability densities functions in Eq. (\ref{eq:probdensityLDP}) to these states
\begin{equation}
\label{eq:densitiesPQ}
P(r)=\frac{e^{-I_P(r)}}{Z_P},\qquad
Q(r)=\frac{e^{-I_Q(r)}}{Z_Q}.
\end{equation}
the large deviation relative entropy is the Kullback-Leibler divergence of these densities \cite{kullback1968information},
\begin{equation}\label{eqn:relative-LDP-def}
D_{\mathrm{LDP}}(P\Vert Q)
=\int_{\mathbb R^d}P(r)\log\frac{P(r)}{Q(r)}\,dr,
\end{equation}
with $\operatorname{supp} P(r) \subseteq \operatorname{supp} Q(r)$, otherwise, we define $D_{\mathrm{LDP}}(P\Vert Q)=+\infty$. So,
\begin{equation}\label{eqn:relative-LDP-rate}
\begin{aligned}
D_{\mathrm{LDP}}(P\Vert Q) &=\braket{I_Q-I_P}_{P}+\log\frac{\mathcal{Z}_Q}{\mathcal{Z}_P}.
\end{aligned}
\end{equation}
The non-negativity of the relative entropy can be seen in a simple way from the inequality $\log u\leq u-1$, indeed,
\begin{align}
-D_{\mathrm{LDP}}(P\Vert Q)
&=\int P\log\frac{Q}{P}\,dr
\leq\int P\left(\frac{Q}{P}-1\right)dz\nonumber\\
&=\int Q\,dr-\int P\,dr=0.
\end{align}
This results shows then that $D_{\mathrm{LDP}}(P\Vert Q)\geq0$, with the equality standing if and only if $P=Q$ almost everywhere. Worth to mention thar this divergence is not symmetric and does not, in general, satisfy the triangle inequality, it measures distinction of the probed distributions, not a metric distance.

\subsection{Conditional Entropy}

Considering a bipartide system $AB$, with coordinates $r_A\in\mathbb R^{d_A}$ and $r_B\in\mathbb R^{d_B}$, and join probability density distribution in Eq. (\ref{eq:probdensityLDP}),
\begin{equation}
P_{AB}(r_A,r_B)=\frac{e^{-I_{AB}(r_A,r_B)}}{Z_{AB}}.
\end{equation}
The relevant marginals are indeed the marginals of the above join distribution, $P_A(r_A)$ and $P_B(r_B)$ by integrating over the coordinates $r_B$ and $r_A$ respectively. In particular, one should not replace $I_{AB}$ with a rate function calculated independently from the reduced Wigner function without first demonstrating that the two constructions coincide. It is useful to rewrite the marginals in Gibbs form. Defining
\begin{align}
I_A^{\mathrm{eff}}(r_A)
&=-\log\int e^{-I_{AB}(r_A,r_B)}\,dr_B,\label{eqn:effective-rate-A}\\
I_B^{\mathrm{eff}}(r_B)
&=-\log\int e^{-I_{AB}(r_A,r_B)}\,dr_A,\label{eqn:effective-rate-B}
\end{align}
then the probability density distributions for the marginals, 
\begin{equation}
P_A(r_A)=\frac{e^{-I_A^{\mathrm{eff}}(r_A)}}{Z_{AB}},
\qquad
P_B(r_B)=\frac{e^{-I_B^{\mathrm{eff}}(r_B)}}{Z_{AB}}.
\end{equation}
Therefore, the conditional density is
\begin{equation}\label{eqn:conditional-density-LDP}
P_{AB}(r_A|r_B) =\frac{P_{AB}(r_A,r_B)}{P_B(r_B)} =\frac{e^{-I_{AB}(r_A,r_B)}}{\int e^{-I_{AB}(r'_A,r_B)}\,dr'_A}.
\end{equation}
The conditional entropy,  
\begin{equation}
S_{\mathrm{LDP}}(A|B=r_B) =-\int P_{AB}(r_A|r_B)\log P_{AB}(r_A|r_B)\,dr_A.
\end{equation}
This quantity still depends on the observed value of $r_B$. The total conditional entropy must be the average of these entropies with respect to the probability of occurrence of each $r_B$, it is straight to check and then the conditional LDP entropy,
\begin{align}
S_{\mathrm{LDP}}(A|B) &=-\iint P_{AB}(r_A,r_B) \log P_{AB}(r_A|r_B)\,dr_A dr_B.\label{eqn:conditional-entropy-LDP}
\end{align}
Considering the definition of the condition probability distribution in Eq.(\ref{eqn:conditional-density-LDP}), we can rewrite the above relation in a closed form
\begin{widetext}
    \begin{align}
S_{\mathrm{LDP}}(A|B) &=-\iint P_{AB}\log P_{AB}\,dr_A dr_B +\iint P_{AB}\log P_B\,dr_A dr_B\nonumber =S_{\mathrm{LDP}}(AB)-S_{\mathrm{LDP}}(B).\label{eqn:conditional-chain-LDP}
\end{align}
\end{widetext}
As it is a differential entropy, $S_{\mathrm{LDP}}(A|B)$ can be negative in some regions.

\subsection{Mutual Information}

We can define mutual information as the relative entropy between the joint density and the product of its marginals:
\begin{equation}\label{eqn:MI-LDP-def}
\mathcal I_{\mathrm{LDP}}(A:B) =D_{\mathrm{LDP}}\big(P_{AB}\Vert P_A P_B\big).
\end{equation}
or even, 
\begin{align}
\mathcal I_{\mathrm{LDP}}(A:B) &=S_{\mathrm{LDP}}(A)+S_{\mathrm{LDP}}(B) -S_{\mathrm{LDP}}(AB)\label{eqn:MI-LDP-entropies} \nonumber\\
&=S_{\mathrm{LDP}}(A)-S_{\mathrm{LDP}}(A|B).
\end{align}
As the relative entropy is nonnegative,
\begin{equation}
\mathcal I_{\mathrm{LDP}}(A:B)\geq0,
\end{equation}
the equality occurs if, and only if,
\begin{equation}
P_{AB}(r_A,r_B)=P_A(r_A)P_B(r_B).
\end{equation}
Thus, $\mathcal I_{\mathrm{LDP}}$ measures all correlations present in the effective density, without automatically separating classical correlations from quantum correlations.

\subsection{Rényi Entropy}

At last, we can study the family of Rényi entropies that can be define from the LDP construction and allowing as well the generalization of the others quantities derived in this section so far. As previously defined
\begin{equation}
\label{eq:densityforRenyi}
P_W(r)=\frac{e^{-I_W(r)}}{ Z_W(1)}, \qquad Z_W(1)=\int e^{-I_W(r)}\,dr,
\end{equation}
(the introduction of the $(1)$ notation in the normalization will become clear in later derivations). Shannon entropy depends on the average of $-\log P_W$. Rényi entropy replaces this average with a family of power moments of the density. For $\alpha>0$ and $\alpha\neq1$, we define
\begin{equation}
S_{\alpha,\mathrm{LDP}}(W) =\frac{1}{1-\alpha} \log\int_{\mathbb R^d}P_W(r)^\alpha\,dr.
\label{eqn:Renyi-LDP-def}
\end{equation}
Powering $P_W$ in Eq. (\ref{eq:densityforRenyi}) to $\alpha$:
\begin{align}
P_W(r)^\alpha =\frac{e^{-\alpha I_W(r)}}{ Z_W(1)^\alpha},
\end{align}
by integrating $P_W^\alpha$
\begin{equation}\label{eqn:Zbeta}
Z_W(\alpha)=\int e^{-\alpha I_W(r)}\,dr, \qquad \alpha>0.
\end{equation}
Hence the normalized density will be
\begin{equation}\label{eqn:escort-LDP}
P_W^{(\alpha)}(r) =\frac{e^{-\alpha I_W(r)}}{ Z_W(\alpha)}.
\end{equation}
For $\alpha > 1$, regions where $I_W$ is large are more strongly suppressed, such that $P_W^{(\alpha)}$ concentrates near the minima of $I_W$. For $0 < \alpha < 1$, the tails receive greater weight. For $\alpha = 1$, we recover $P_W^{(1)} = P_W$. Formally, $\alpha$ plays the role of an inverse temperature applied to the ``effective energy'' $I_W$ (as in standard statistical mechanics), without the need to assign it a physical temperature. So, the moments of $P_W(r)$,
\begin{align}
\int P_W(r)^\alpha\,dr &=\int\left(\frac{e^{-I_W(r)}}{ Z_W(1)}\right)^\alpha dr\nonumber =\frac{ Z_W(\alpha)}{ Z_W(1)^\alpha}.
\end{align}
Substituting into Eq. (\ref{eqn:Renyi-LDP-def})
\begin{equation}
S_{\alpha,\mathrm{LDP}}(W) = \frac{\log Z_W(\alpha) -\alpha\log Z_W(1)}{1-\alpha}.
\end{equation}
Next, as the limit of $\alpha \rightarrow 1$ in the Rényi entropy is often studied, we will start by defining the first and second derivatives of the cumulant-generating function, i.e., the derivatives of $\log Z_W(\alpha)$. The derivatives of the partition function are defined as
\begin{align}
 Z_W'(\alpha) &=-\int I_W(r)e^{-\alpha I_W(r)}\,dr,\\
 Z_W''(\alpha) &=\int I_W(r)^2e^{-\alpha I_W(r)}\,dr,
\end{align}
the first derivative of the cumulant-generating function,
\begin{align}
\frac{d}{d\alpha}\log Z_W(\alpha) &=\frac{ Z_W'(\alpha)}{ Z_W(\alpha)}=-\braket{I_W}_{\alpha},
\end{align}
while for the second derivative
\begin{equation}\label{eqn:Zalpha-variance}
\frac{d^2}{d\alpha^2}\log Z_W(\alpha) =\braket{I_W^2}_{\alpha}-\braket{I_W}_{\alpha}^2
=\operatorname{Var}_{\alpha}(I_W)\geq0.
\end{equation}
We see than that the first derivative provides the mean of the rate function and the second its variance (fluctuation). Next, taking the limit $\alpha\to1$ in Eq. \eqref{eqn:Renyi-LDP-def}, we see both numerator and denominator goes to zero, from the L'Hospital formula, 
\begin{align}
\lim_{\alpha\to1}S_{\alpha,\mathrm{LDP}}(W) =\braket{I_W}_{P_W}+\log Z_W(1) =S_{\mathrm{LDP}}(W).
\end{align}
Therefore, the family of Rényi entropies is a continuum deformation of the standard large deviation entropy $S_{\text{LDP}}$. Different values of $\alpha$ probe different regions of the rate function. We can define then the Rényi entropic form for both the relative entropy in Eq.(\ref{eqn:relative-LDP-rate}) as well as the mutual information in Eq. (\ref{eqn:MI-LDP-def}). The relative Rényi entropy is defined as
\begin{equation}
D_{\alpha,\mathrm{LDP}}(P\Vert Q) =\frac{1}{\alpha-1} \log\int P(r)^\alpha Q(r)^{1-\alpha}\,dr.
\end{equation}
Following the definitions for $P(r)$ and $Q(r)$ in Eq. (\ref{eq:densitiesPQ}), then
\begin{equation}\label{eqn:relative-Renyi-rate}
D_{\alpha,\mathrm{LDP}}(P\Vert Q) =\frac{1}{\alpha-1}\log \frac{\displaystyle\int
e^{-\alpha I_P(r)-(1-\alpha)I_Q(r)}dr} {Z_P^\alpha Z_Q^{1-\alpha}}.
\end{equation}
In the limit of $\alpha\to1$, we do recover $D_{\mathrm{LDP}}$. The Rényi mutual information could be then defined
\begin{equation}
\mathcal I_{\alpha,\mathrm{LDP}}(A:B) =D_{\alpha,\mathrm{LDP}} \big(P_{AB}\Vert P_AP_B\big).
\end{equation}
In general, one should not simply replace it with $S_\alpha(A)+S_\alpha(B)-S_\alpha(AB)$, since this combination does not possess all the properties of mutual information for any $\alpha$.

\subsubsection{Rényi-Wigner Second Order Entropy}

The family $S_{\alpha,\mathrm{LDP}}$ should not be confused with the quantum Rényi entropy obtained directly from the state purity \cite{PRA2021, PRA2023}. These are two different constructions:
\begin{align}
S_{2,\mathrm{LDP}}(W) &=-\log\int P_W(r)^2\,dr,\\
S_2^{(W)}(\rho) &=-\log\left[(2\pi)^n\int W_\rho(r)^2\,d^{2n}r\right].
\end{align}
The first uses the auxiliary positive density $P_W=e^{-I_W}/\mathcal{Z}_W$ while the second directly uses the square of the Wigner distribution and coincides with the quantum entropy of Rényi of order two. For a density operator $\rho$, in the convention used in this chapter, the Wigner distribution reads
\begin{equation}
W_\rho(x,p)=\frac{1}{2\pi} \int_{-\infty}^{\infty} e^{ipy} \bra{x-y/2}\rho\ket{x+y/2}\,dy.
\end{equation}
One can check the following purity relation, 
\begin{equation}\label{eqn:Wigner-purity}
(2\pi)^n\int_{\mathbb R^{2n}}W_\rho(r)^2\,d^{2n}r =\operatorname{Tr}(\rho^2),
\end{equation}
with $0<\operatorname{Tr}(\rho^2)\leq1$ which is equal for pure states and less than one for mixed states. Then,
\begin{widetext}
    \begin{equation}\label{eqn:Renyi2-Wigner}
S_2^{(W)}(\rho) =-\log\left[(2\pi)^n\int W_\rho(r)^2\,d^{2n}r\right] =-\log\operatorname{Tr}(\rho^2)=S_2(\rho).
\end{equation}
\end{widetext}
This identity remains valid when $W_\rho$ is negative, since its square appears. This quantity is interesting because it can be used to measure entanglement in pure states. Consider a globally pure bipartite state in its Schmidt decomposition,
\begin{equation}
\ket{\Psi_{AB}}=\sum_j\sqrt{\lambda_j} \ket{a_j}\ket{b_j}, \qquad \lambda_j\geq0,
\qquad \sum_j\lambda_j=1,
\end{equation}
the reduces state of $A$ is
\begin{align}
\rho_A &=\operatorname{Tr}_B\ket{\Psi_{AB}}\bra{\Psi_{AB}}=\sum_j\lambda_j\ket{a_j}\bra{a_j},
\end{align}
and has purity $\operatorname{Tr}(\rho_A^2)=\sum_j\lambda_j^2$. Thus,
\begin{equation}
S_2^{(W)}(\rho_A) =-\log\sum_j\lambda_j^2.
\end{equation}
If the state is a product state, there is only one non-zero Schmidt coefficient, $\lambda_1=1$, and $S_2^{(W)}(\rho_A)=0$. If the state is entangled, there are at least two non-zero coefficients, $\sum_j\lambda_j^2<1$, and $S_2^{(W)}(\rho_A)>0$. Since $\rho_A$ and $\rho_B$ have the same Schmidt coefficients, $S_2^{(W)}(\rho_A)=S_2^{(W)}(\rho_B)$. As a sort example, let us consider the vacuum state of the 1D harmonic oscillator. For one mode in the vacuum state,
\begin{equation}
W_0(x,p)=\frac{1}{\pi}e^{-x^2-p^2}, \,\,\,\, \int W_0(x,p)^2\,dx\,dp =\frac{1}{2\pi}.
\end{equation}
Consequently,
\begin{equation}
(2\pi)\int W_0^2=1, \qquad S_2^{(W)}(\ket0\bra0)=0,
\end{equation}
as we expect for a pure state. However, as in this gaussian scenario $P_W=W_0$,
\begin{equation}
S_{2,\mathrm{LDP}}(W_0) =-\log\int W_0^2 =\log(2\pi).
\end{equation}
The example clarifies the difference: $S_{2,\mathrm{LDP}}$ is a differential entropy of the effective density, whereas $S_2^{(W)}$ contains the measure factor $(2\pi)^n$ required to match the quantum purity. For a positive Gaussian Wigner function, the two differ by the constant
\begin{equation}
S_2^{(W)}(\rho)=S_{2,\mathrm{LDP}}(W)-n\log(2\pi).
\end{equation}
For general orders, one should not identify $\int W^\alpha$ with $\operatorname{Tr}(\rho^\alpha)$: if $\alpha$ is not an integer, the negativity of $W$ can even yield complex values.

\section{Entropic Uncertainty Relations and Quantum Correlations}
\label{sec:Entropic_Uncertainty_Quantum_Correlations}
\subsection{Entropic Uncertainty Relations}

\label{sec:EUR-summary-LDP}

Entropic uncertainty relations quantify the impossibility of simultaneously concentrating the distributions of two conjugate quadratures, for the results present in this section, we are adopting the same convention as in \cite{FloerchingerHaasMueller2021}. For a position variable $x$ and its conjugate moment $p$, let
\begin{equation}
f(x)=\langle x|\rho|x\rangle, \qquad g(p)=\langle p|\rho|p\rangle,
\end{equation}
and define the differential entropy
\begin{equation}
S_{\text{sh}}(f)=-\int_{\mathbb R}f(x)\log f(x)\,dx, \,\, S_{\text{sh}}(g)=-\int_{\mathbb R}g(p)\log g(p)\,dp.
\end{equation}
In units $\hbar=1$, the Bia\l ynicki-Birula-Mycielski relation \cite{BialynickiBirulaMycielski1975} (defined from the marginals) is
\begin{equation}
\label{eqn:EUR-BBM-summary}
S_{\text{sh}}(f)+S_{\text{sh}}(g)\geq\log(e\pi).
\end{equation}
The Frank-Lieb inequality \cite{FrankLieb2012} (defined by the marginals and the Shannon entropy of the complete state) also incorporates the quantum mixing of the state,
\begin{equation}\label{eqn:EUR-FL-summary}
S_{\text{sh}}(f)+S_{\text{sh}}(g)\geq\log(2\pi)+S(\rho),
\end{equation}
while the Wehrl-Lieb inequality \cite{Lieb1978, WehrlLieb84} (defined only from Wehrl entropy) is
\begin{equation}\label{eqn:EUR-WL-summary}
S_W(\rho)\geq1.
\end{equation}
To compare how close each relationship is to equality, following the convention adopted in \cite{FloerchingerHaasMueller2021}, we set the lower bound of the inequalities as $\log(e\pi)$. Therefore
\begin{equation}
\left.
\begin{aligned}
\mathcal U_{\mathrm{WL}} &= S_W(\rho)+\log\pi,\\
\mathcal U_{\mathrm{BBM}} &= S_{\text{sh}}(f)+S_{\text{sh}}(g),\\
\mathcal U_{\mathrm{FL}} &= S_{\text{sh}}(f)+S_{\text{sh}}(g)-S(\rho)+1-\log2,
\end{aligned}
\right\}
\geq \mathcal \log(e\pi).
\label{eqn:EUR-common-scale}
\end{equation}
For the LDP entropy we need to have its measurement convention fixed before being compared with the others. In the convention used in this work, the LDP density constructed from the vacuum yields as well $S_{\mathrm{LDP}}^{(W)}(\ket0)=\log(e\pi)$. In the Gaussian sector, LDP relation can be demonstrated, otherwise it will be presented as a conjectured, verified in the family of states presented in this paper. A general proof would require showing that the effective vacuum density minimizes the LDP entropy over the considered class of states. The positivity of $P_{\mathrm{LDP}}$ guarantees that its entropy is real, but does not alone imply the $\mathcal U_{\mathrm{LDP}-W} \ge \log (e \pi$) limit.

\begin{figure}
    \centering
    \includegraphics[width=0.9\linewidth]{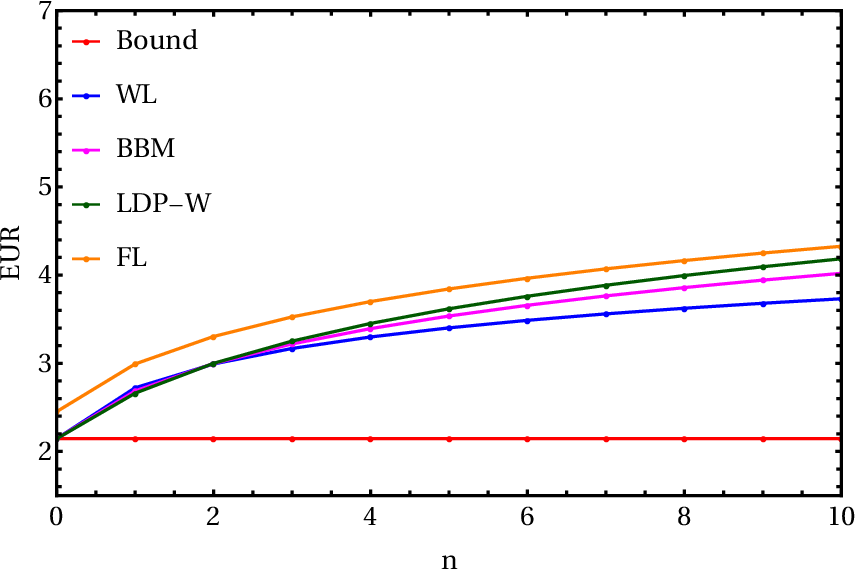}
    \includegraphics[width=0.91\linewidth]{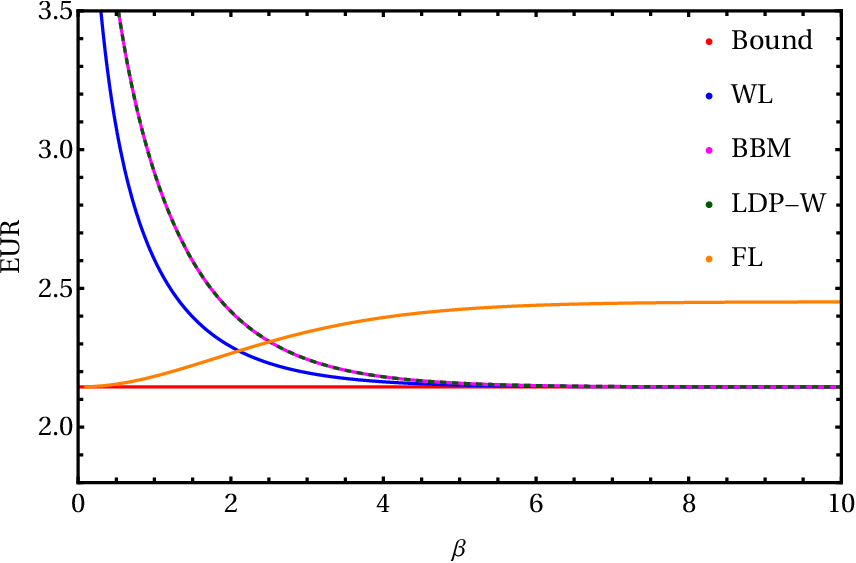}
    \caption{In this figure, we present a comparison between different entropic uncertainty relations. The top panel shows the analysis for pure states of a harmonic oscillator as a function of the energy levels $n$. Here, we denote the uncertainty relation derived in this work as $\text{LDP-W}$. The bottom panel displays the same set of uncertainty relations, but for mixed states as a function of the inverse temperature, $\beta$.}
    \label{fig:EUR1}
\end{figure}

The upper panel of Figure \ref{fig:EUR1} shows how WL, BBM, and the LDP constructs respond differently to increasing $n$. In particular, the LDP construction is shown to be more tight than the FL relation, and is similar for the first states of the harmonic oscillator.  

For the thermal state, the behavior is complementary, as can be seen in the lower panel of Figure \ref{fig:EUR1}.  In the high-temperature regime, $\beta\to0$, the $S(\rho)$ contribution makes the Frank-Lieb relation become especially close to equality, while relations based only on phase dispersion increase. In particular we that the LDP prescription matches the BBM relation. The uncertainty relation for the Wigner entropy of Gaussian states have also been explored (see \cite{Van_Herstraeten_2025} for a recent analysis and references therein), where it matches the lower bound of the LDP formulation, i.e., $\ln \pi + 1$. 

\subsection{Mutual Information and Entanglement Witness}

As also pointed out in \cite{FloerchingerHaasMueller2021}, the mutual information can be used as a witnesses of entanglement, we then check how this feature is fulfilled for the large deviation entropy. Consider now a bipartite state $\rho_{AB}$, although the Wigner function can take negative values, the large deviation principle shows that the generating function is positive, convex, and analytic. 

\begin{prop}\label{prop:pure-negative-Wigner-witness}
For a globally pure bipartite state,
\begin{equation}
\mathcal I_{\mathrm{LDP}}^{(W)}(A:B)=0 \quad\Longleftrightarrow\quad \rho_{AB}=\rho_A\otimes\rho_B.
\end{equation}
Consequently,
\begin{equation}
\mathcal I_{\mathrm{LDP}}^{(W)}(A:B)>0 \quad\Longleftrightarrow\quad
\rho_{AB}\ \text{is entangled},
\end{equation}
even when $W_{AB}$ has negative regions.
\end{prop}
The proof can be seen in detail in Appendix \ref{app_proof_prep_witness}. For mixed states, the result does not hold: a classically correlated separable state can produce $\mathcal I_{\mathrm{LDP}}^{(W)}>0$ or even quantum correlation beyond entanglement.
\section{Discussion and Outlook}
\label{sec:Discussion_Outlook}

In this work, we primarily address the connection between large deviation theory and the Wigner distribution. The objective of this study was to construct a framework in which the Shannon entropy of a phase-space distribution could be well-defined without the necessity of a smearing procedure of the phase-space. By defining a rate function for the Wigner distribution, we were able to construct a probability density for which an entropy could be established, which we called \textit{large deviation principle entropy}. For the case of the harmonic oscillator, we show that this entropy is greater than the Wehrl entropy. Although this might seem counterintuitive, given that the Wehrl entropy relies on a Gaussian smoothing of the Wigner distribution, we argue that our entropy formulation effectively captures all the correlations and structures present in the Wigner distribution, features that are typically spread when deriving the Husimi distribution. As a further motivation for this study, we derive other fundamental relations for $S_{\text{LDP}}$, such as relative entropy, conditional entropy, mutual information, and Rényi entropy. Mutual information, in particular, is of interest as it can probe and provide a lower bound on the entanglement of a system in the case of pure states. Furthermore, the entropic uncertainty relation for $S_{\text{LDP}}$ yields a lower bound for the vacuum state consistent with those previously discussed in the literature.

The formalism developed in this work could be suitable to study some class of problem such as thermalization in closed quantum systems, entropy production in open quantum systems, quantum correlations beyond entanglement, non-Gaussianity among others. In particular, since the large deviation principle entropy can be defined on a local level it could serve as an initial condition for hydrodynamics in a strongly fluctuating quantum regime, such as hadronic collisions and strongly correlated cold atom systems, in a manner analogous to \cite{RabeloSoares25}.

In the context of thermalization, the Eigenstate Thermalization Hypothesis (ETH) (for a general review, see \cite{Deutsch_2018}) provides a framework for understanding how isolated quantum systems, composed of many degrees of freedom and exhibiting non-integrable dynamics, can display thermal behavior in the absence of an external heat bath. Motivated by ideas from quantum chaos and Berry's conjecture \cite{berry1,berry2,berry3}, the ETH states that the expectation values of local observables in individual energy eigenstates are, under appropriate conditions, equivalent to those obtained from a microcanonical ensemble at the corresponding energy. In other words, individual eigenstates themselves encode the statistical properties associated with thermal equilibrium.

Quantum information theory has provided tools for investigating the structure of eigenstates, the thermalization of subsystems, and quantum correlations in ETH scenario. Entanglement entropy, for instance, is a well-established approach in this field. Although the global isolated system corresponds to a pure state, meaning it has a von Neumann entropy equal zero, one can examine the entanglement between subsystems, thereby gaining insight into how thermodynamic entropy emerges from the structure of the eigenstates \cite{Liu18,Srednicki19,Grover19}. From this perspective, the entropy framework developed in this work provides a complementary phase-space approach to investigating the statistical structure of quantum states. As the formulation of the entropy lies on the intersection of the Wigner distributions and large deviation theory, its microcanonical-type interpretation suggests a natural avenue for investigating the statistical properties of quantum energy eigenstates and their relation to thermalization. Establishing how this entropy behaves across eigenstates within a narrow energy window, and how its predictions compare with those of the corresponding microcanonical description, could provide a useful phase-space perspective on the ETH.

On the other hand, in the context of open quantum systems, Wigner distribution has been used to reformulate dissipative dynamics and entropy production. In this scenario, the Wigner relative entropy provides the entropy production of the system \cite{Santos_2017,Landi21}, which is well defined by positive Wigner distribution. In principle, these ideas could be easily extended to non-Gaussian Wigner distributions by considering a formulation based on large deviation theory. In addition it also provides a suitable framework to study the time dependent formulation of the $S_{\text{LDP}}$ and its dynamics.

The last result present in this work was connection of the mutual information as a witness of entanglement for pure states when $\mathcal I_{\mathrm{LDP}}^{(W)}(A:B)>0$. Conversely, this condition can also be fulfilled for mixed states, but the conclusion is not straight, for mixed states, there are other types of quantum correlations beyond the entanglement. For this correlation one can have, for instance, quantum discord \cite{Discord2001,Vedral2001} where this quantity can be defined as the difference between mutual information and classical correlations. In addition, in other scenarios, relative entropy was used to probe the non-Gaussianity of the system \cite{Paris2008}. In this sense, these two examples: quantum discord and non-Gaussianity, which have previously been explored using quantities such as mutual information, classical correlation, and relative entropy, could also be studied through the quantities defined in the present work, this would be an interesting topic for future work.

Yet on the result from the Preposition \ref{prop:pure-negative-Wigner-witness}, would also be interesting to check whether $\mathcal I_{\mathrm{LDP}}^{(W)}(A:B)$ fulfills the requirements of an entanglement monotone \cite{Vidal_2000} as invariance over local unitary transformations and monotonicity over LOCC (Local Operations and Classical Communication). This was not checked in the present work, and the result is that at least $\mathcal I_{\mathrm{LDP}}^{(W)}(A:B)$ is good witnesses of entanglement for pure states.

\section{Conclusions}
\label{sec:conclusions}
In this work, we propose a formal framework for calculating the entropy of a Wigner distribution, applicable to both Gaussian and non-Gaussian states. To define this quantity, we employ the theory of large deviations, which allows for the construction of a positive valued function, the cumulant-generating function, even when the Wigner distribution is not positive definite. We thus establish a formal procedure for defining these quantities and, consequently, the entropy. In this context, we perform calculations for a one-dimensional harmonic oscillator, considering both mixed and pure states. Furthermore, we derive additional quantities, such as relative entropy, conditional entropy, mutual information, and the Rényi entropy family, within the framework of large-deviation entropy. We also examine how this entropy fits into the context of entropic uncertainty relations comparing with results already present in the literature. Finally, we discuss the general results obtained in this work and explored potential scenarios in which this entropy might be useful and applicable. 

\section*{Acknowledgments}

G.R.S. thanks Leonardo AM Souza for a fruitful discussion on aspects of quantum information theory. G.R.S. acknowledge the hospitality at IFJ PAN where part of this work was conducted. G.R.S. is supported by the CAPES doctoral fellowship 88887.005836/2024-00 and 88881.263438/2026-01. Y.K. was supported by FAPESP 2025/22647-3 Jovem Pesquisador award. G.T. thanks Bolsa de produtividade CNPQ 305731/2023-8 and FAPESP 2023/06278-2, as well as FAPESP temático 2023/13749-1 for support. 
\bibliography{ref}

\newpage
\newpage
\onecolumngrid 
\appendix 

\section{Moments of the Wigner Distribution and the McCoy's formula}
\label{app_moments_charc_function}
In this appendix we derive the McCoy's formula that provides a general expression of de (n,m)-moment of the Wigner distribution. Setting $\hbar=1$, following the Wigner distribution defined in Eq.\eqref{eq:wignerdef}. Here, $\psi(x)$ is the position function and $\widehat{\psi}(p)={1 \over \sqrt{2\pi}}\int\limits_{-\infty}^\infty  \psi(x) e^{-ipx}\,dx$ is the momentum. For any $m,n \in \mathbb{N}$, the corresponding $(n,m)$-moment of $W(x,p)$,
\begin{align}\label{eqn:covWnm}
\iint\limits_{\mathbb{R}^2} \! x^n p^m \, W(x,p) \,dx\, dp 
&={1 \over 2\pi} \iiint\limits_{\mathbb{R}^3}\! x^n p^m \, \psi^*(x+y/2) \, \psi(x-y/2) e^{ipy}\,dy  \,dx\, dp  \nonumber \\
&={(-i)^m \over 2\pi}  \iiint\limits_{\mathbb{R}^3}\! x^n \, \psi^*(x+y/2) \, \psi(x-y/2) {\partial^m \over \partial y^m}e^{ipy}\,dy  \,dx\, dp \nonumber \\
&={i^m \over 2\pi}  \iiint\limits_{\mathbb{R}^3}\! x^n \,\left[{\partial^m \over \partial y^m} \Big(\psi^*(x+y/2) \, \psi(x-y/2)\Big)\right] e^{ipy}\,dy  \,dx\, dp,
\end{align}
where
\be\label{eqn:dym}
{\partial^m \over \partial y^m} \Big(\psi^*(x+y/2) \, \psi(x-y/2)\Big)
={1 \over 2^m}\sum\limits_{j=0}^m {m \choose j}(-1)^{m-j}\left(\psi^{(j)}(x+y/2)\right)^* \, \psi^{(m-j)}(x-y/2).
\ee

\medskip
\noindent
Next, substituting \eqref{eqn:dym} into \eqref{eqn:covWnm} and letting $u=x+y/2$ and $v=x-y/2$ yields
\begin{align}\label{eqn:covWnm2}
 \iint\limits_{\mathbb{R}^2} \! x^n p^m \, W(x,p) \,dx\, dp \nonumber 
&={1 \over 2^{n+m}}\sum\limits_{j=0}^m {m \choose j}{(-1)^{m-j} i^m \over 2\pi}  \iiint\limits_{\mathbb{R}^3}\! (u+v)^n \,\left(\psi^{(j)}(u)e^{-ipu}\right)^* \, \psi^{(m-j)}(v) e^{-ipv} \,du  \,dv\, dp \nonumber \\
&={1 \over 2^{n+m}}\sum\limits_{j=0}^m {m \choose j}{(-1)^{m-j} i^m \over 2\pi}  \sum\limits_{\ell=0}^n {n \choose \ell}\!\!\iiint\limits_{\mathbb{R}^3}\! \left(u^\ell \psi^{(j)}(u)e^{-ipu}\right)^* v^{n-\ell} \psi^{(m-j)}(v) e^{-ipv} du  \,dv\, dp,
\end{align}
where for integers $\ell$ and $n$ the Fourier transform
$$\int\limits_{-\infty}^\infty x^\ell \psi^{(j)}(x)e^{-ipx}\,dx=\sqrt{2 \pi} \,i^{\ell+j} {d^\ell \over dp^\ell}\big(p^j \widehat{\psi}(p) \big)
=\sqrt{2 \pi} \,i^j {\sf x}^\ell p^j \widehat{\psi}(p),$$
as ${\sf x}=i{d \over dp}$ in the momentum space. Therefore,
\begin{align}\label{eqn:covWnm3}
 \iint\limits_{\mathbb{R}^2}  x^n p^m \, W(x,p) \,dx\, dp \nonumber
&={1 \over 2^{n+m}}\sum\limits_{j=0}^m {m \choose j}(-1)^{m-j} i^m  \sum\limits_{\ell=0}^n {n \choose \ell}\!\!\int\limits_{-\infty}^\infty \! \left(i^j {\sf x}^\ell p^j \widehat{\psi}(p) \right)^* i^{m-j} {\sf x}^{n-\ell} p^{m-j} \widehat{\psi}(p)  \,dp \nonumber \\
&={1 \over 2^{n+m}}\sum\limits_{j=0}^m {m \choose j}  \sum\limits_{\ell=0}^n {n \choose \ell}\!\!\int\limits_{-\infty}^\infty \! \left({\sf x}^\ell p^j \widehat{\psi}(p) \right)^* {\sf x}^{n-\ell} p^{m-j} \widehat{\psi}(p)  \,dp \nonumber \\
&={1 \over 2^{n+m}}\sum\limits_{j=0}^m {m \choose j}  \sum\limits_{\ell=0}^n {n \choose \ell} \langle {\sf p}^i {\sf x}^n {\sf p}^i \rangle ={1 \over 2^m}\sum\limits_{j=0}^m {m \choose j} \langle {\sf p}^j {\sf x}^n {\sf p}^{m-j} \rangle,
\end{align}
as $\sum\limits_{\ell=0}^n {n \choose \ell}=2^n$.
Thus, we confirm the following symmetrization formula for the $(n,m)$-moment of $W(x,p)$,
\be\label{eqn:McCoy}
 \iint\limits_{\mathbb{R}^2} x^n p^m \, W(x,p) \,dx\, dp
={1 \over 2^m}\sum\limits_{j=0}^m {m \choose j} \langle {\sf p}^j {\sf x}^n {\sf p}^{m-j} \rangle,
\ee
where the symmetrized expression in the right hand side of \eqref{eqn:McCoy}, before taking the expectation, is known as {\it McCoy's formula}.

\section{Law of Large Numbers and the Central Limit Theorem}
\label{app_central_limit_theorem}
In this section we introduce briefly the Law of Large Numbers and the Central Limit Theorem results. By the analogy of the Fourier transform proof of the Weak Law of Large Numbers (WLLN), we consider the limit
$$\lim\limits_{n \to \infty}\Big(\widehat{\widehat{W}}\big(t/n,\, k/n \big)\Big)^n.$$
The Taylor expansion in Eq. \eqref{eqn:WhhMc} yields
\be\label{eqn:WhhnMc}
\widehat{{\widehat{W}}}\big(t/n,\, k/n \big)=\langle e^{-ik{\sf p} \over 2n} e^{-it{\sf x} \over n} e^{-ik{\sf p} \over 2n} \rangle 
=1-i{t \over n}\langle{\sf x} \rangle-i{k \over n}\langle{\sf p} \rangle +O\left({1 \over n^2}\right).
\ee
Thus, we obtain the following version of the law of large numbers for the Wigner quasiprobability distribution:
\begin{align}\label{eqn:Wlln}
\lim\limits_{n \to \infty}\Big(\widehat{\widehat{W}}\big(t/n,\, k/n \big)\Big)^n
&=\lim\limits_{n \to \infty}\left(1-i{t\over n}\langle{\sf x} \rangle -i{k \over n}\langle{\sf p} \rangle+O\left({1 \over n^2}\right) \right)^n
=e^{-it\langle{\sf x} \rangle-ik\langle{\sf p} \rangle} \nonumber \\
&=\mathcal{F}_{x,p}[\delta_{\langle{\sf x} \rangle,\langle{\sf p} \rangle}](t,k),
\end{align}
where
$\delta_{\langle{\sf x} \rangle,\langle{\sf p} \rangle}$ denotes the delta point mass concentrated at $(\langle{\sf x} \rangle,\langle{\sf p} \rangle)$ in the position-momentum space.
Observe that in the case when $W(x,p)$ is nonnegative, \eqref{eqn:Wlln} is equivalent to the standard weak law of large numbers.
Following the approach used in the proof of the classical central limit theorem (CLT), we consider the limit
$$\lim\limits_{n \to \infty}\left(\widehat{\widehat{W}}\left({t \over \sqrt{n}},\, {k \over \sqrt{n}} \right) e^{i{t\over \sqrt{n}}\langle{\sf x} \rangle +i{k \over \sqrt{n}}\langle{\sf p} \rangle}\right)^n,$$
where Eq. \eqref{eqn:WhhMc} yields the following Taylor approximation
\be\label{eqn:hhW2n2n}
\widehat{\widehat{W}}\left({t \over \sqrt{n}},\, {k \over \sqrt{n}}\right) =1-i{t\over \sqrt{n}}\langle{\sf x} \rangle -i{k \over \sqrt{n}}\langle{\sf p} \rangle -{t^2 \over 2n}\langle{\sf x}^2 \rangle -{tk \over 2n}(\langle{\sf xp} \rangle+\langle{\sf px} \rangle) -{k^2 \over 2n}\langle{\sf p}^2 \rangle +O\left({1 \over n^{3/2}}\right).
\ee

\medskip
\noindent
Next, we combine \eqref{eqn:hhW2n2n} with $\,\log(1-h)=-h-{1 \over 2}h^2+O(h^3)$, obtaining
\begin{align}\label{eqn:loghhW2n2n}
n\, \log &\left(\widehat{\widehat{W}}\left({t \over \sqrt{n}},\, {k \over \sqrt{n}} \right) e^{i{t\over \sqrt{n}}\langle{\sf x} \rangle +i{k \over \sqrt{n}}\langle{\sf p} \rangle}\right) \nonumber \\
&=n\,\bigg[ \log\left(1-i{t\over \sqrt{n}}\langle{\sf x} \rangle -i{k \over \sqrt{n}}\langle{\sf p} \rangle -{t^2 \over 2n}\langle{\sf x}^2 \rangle -{tk \over 2n}(\langle{\sf xp} \rangle+\langle{\sf px} \rangle) -{k^2 \over 2n}\langle{\sf p}^2 \rangle +O\left({1 \over n^{3/2}}\right)\right) 
+i{t\over \sqrt{n}}\mu +i{k \over \sqrt{n}}\gamma. \bigg] \nonumber \\
&=-{t^2 \over 2}\sigma_{\sf x}^2 -tk c_{\sf xp} -{k^2 \over 2} \sigma_{\sf p}^2+O\left({1 \over \sqrt{n}}\right),
\end{align}
where $\sigma_{\sf x}^2=\langle{\sf x}^2 \rangle -(\langle{\sf x} \rangle)^2$, $~\sigma_{\sf p}^2=\langle{\sf p}^2 \rangle -(\langle{\sf p} \rangle)^2$, and 
{\it quasicovariance} 
$$c_{\sf xp}={\langle{\sf xp} \rangle+\langle{\sf px}\rangle \over 2}-\langle{\sf x} \rangle\langle{\sf p} \rangle
= \iint\limits_{\mathbb{R}^2} \! \big(x-\langle{\sf x} \rangle\big) \big(p-\langle{\sf p} \rangle\big) \, W(x,p) \,dx\, dp.$$ 
 Eq. \eqref{eqn:loghhW2n2n} can be rewritten in the matrix form as follows
\be\label{eqn:loghhW2n2nM}
n\, \log\left(\widehat{\widehat{W}}\left({t \over \sqrt{n}},\, {k \over \sqrt{n}} \right) e^{i{t\over \sqrt{n}}\mu +i{k \over \sqrt{n}}\gamma}\right)
=-{1 \over 2}\big(t,~k\big)\left(\begin{array}{cc}\sigma_{\sf x}^2 & c_{\sf xp} \\c_{\sf xp} &  \sigma_{\sf p}^2\end{array}\right)
\left(\!\begin{array}{c}t \\k\end{array}\!\right) +O\left({1 \over \sqrt{n}}\right).
\ee
Thus, we obtained the following version of central limit theorem
\be\label{eqn:clt}
\lim\limits_{n \to \infty}\left(\widehat{\widehat{W}}\left({t \over \sqrt{n}},\, {k \over \sqrt{n}} \right) e^{i{t\over \sqrt{n}}\langle{\sf x} \rangle +i{k \over \sqrt{n}}\langle{\sf p} \rangle}\right)^n
=\exp\left\{-{1 \over 2}\big(t,~k\big)\Sigma \left(\!\!\begin{array}{c}t \\k\end{array}\!\!\right)\right\}, 
\ee
where
\be\label{eqn:Sigma}
\Sigma=\left(\begin{array}{cc}\sigma_{\sf x}^2 & c_{\sf xp} \\c_{\sf xp} &  \sigma_{\sf p}^2\end{array}\right),
\ee
is the {\it covariance matrix}, easily verified to be positive definite. Indeed, by the Cauchy-Bunyakovsky-Schwarz inequality,
$$\sigma_{\sf x}\sigma_{\sf p} \geq \big|\langle{\sf xp} \rangle-\langle{\sf x} \rangle\langle{\sf p} \rangle\big|>c_{\sf xp},$$
as
$\langle  {\sf xp} \rangle={\langle{\sf xp} \rangle+\langle{\sf px}\rangle \over 2}+i{\hbar \over 2}$ and $\langle  {\sf px} \rangle=\langle{\sf xp} \rangle^*$.
Observe that for nonnegative $W(x,p)$, equation \eqref{eqn:clt} yields the classical central limit theorem.
\section{Proof of Lemma I}
\label{app_cumulant_convexity}
\begin{proof}
In order to show that $\log M_W(t,k)$ is convex, we need to establish that the determinant of its Hessian $\mathcal{H}_{t,k}={\rm Hess}\big(\log M_W(t,k)\big)$ is nonnegative .

\medskip
\noindent
For a given pair $t,k \in \mathbb{R}$, define the following bracket relation
$$\ll f({\sf x},{\sf p}) \gg={1 \over M_W(t,k)}\langle e^{k{\sf p}/2} e^{t{\sf x}/2} f({\sf x},{\sf p}) e^{t{\sf x}/2} e^{k{\sf p}/2} \rangle.$$
The Hessian $\mathcal{H}_{t,k}$ can be expressed via $\ll \cdot \gg$ brackets using the following commutation relations
\be\label{eqn:com}
{\sf p}^ne^{t{\sf x}/2}=e^{t{\sf x}/2}\big(-it/2+{\sf p}\big)^n \quad \text{ and } \quad
e^{t{\sf x}/2}{\sf p}^n=\big(it/2+{\sf p}\big)^ne^{t{\sf x}/2} ,
\ee
for any $n \in \mathbb{N}$ as ${\sf p}=-i{d \over dx}$. Indeed, Eq. \eqref{eqn:com} yields
\begin{align}\label{eqn:Htt}
{\partial^2 \over \partial t^2} \log M_W(t,k) 
&={\langle e^{k{\sf p}/2} {\sf x}^2 e^{t{\sf x}} e^{k{\sf p}/2} \rangle \over M_W(t,k)}
-{\big(\langle e^{k{\sf p}/2} {\sf x} e^{t{\sf x}} e^{k{\sf p}/2} \rangle \big)^2 \over M_W^2(t,k)}  \nonumber \\
&=\ll {\sf x}^2 \gg -(\ll {\sf x} \gg)^2,
\end{align}
while,
\begin{align}\label{eqn:Htk}
{\partial^2 \over \partial t \partial k} \log M_W(t,k) 
&={\langle e^{k{\sf p}/2} {\sf px} e^{t{\sf x}} e^{k{\sf p}/2} \rangle+\langle e^{k{\sf p}/2} e^{t{\sf x}} {\sf xp} e^{k{\sf p}/2} \rangle \over 2 M_W(t,k)}  -{\big(\langle e^{k{\sf p}/2} {\sf p} e^{t{\sf x}} e^{k{\sf p}/2} \rangle+\langle e^{k{\sf p}/2} e^{t{\sf x}} {\sf p} e^{k{\sf p}/2} \rangle \big) \langle e^{k{\sf p}/2} {\sf x} e^{t{\sf x}} e^{k{\sf p}/2} \rangle \over 2 M_W^2(t,k)} \nonumber \\
&={1 \over 2}\ll{\sf xp}+{\sf px} \gg  -\ll {\sf x} \gg \ll {\sf p} \gg,
\end{align}
and
\begin{align}\label{eqn:Hkk}
{\partial^2 \over \partial k^2} \log M_W(t,k) 
&={\langle e^{k{\sf p}/2} {\sf p}^2 e^{t{\sf x}} e^{k{\sf p}/2} \rangle+2\langle e^{k{\sf p}/2} {\sf p}e^{t{\sf x}}{\sf p} e^{k{\sf p}/2} \rangle+\langle e^{k{\sf p}/2} e^{t{\sf x}} {\sf p}^2 e^{k{\sf p}/2} \rangle \over 4 M_W(t,k)} -{\big(\langle e^{k{\sf p}/2} {\sf p} e^{t{\sf x}} e^{k{\sf p}/2} \rangle+\langle e^{k{\sf p}/2} e^{t{\sf x}} {\sf p} e^{k{\sf p}/2} \rangle \big)^2 \over 4 M_W^2(t,k)} \nonumber \\
&=\ll {\sf p}^2 \gg -(\ll {\sf p} \gg)^2.
\end{align}
Following that,
$\,\big(\!-it/2+{\sf p}\big)+\big(it/2+{\sf p}\big)=2{\sf p},\,$
$~~\big(\!-it/2+{\sf p}\big){\sf x}+{\sf x}\big(it/2+{\sf p}\big)={\sf xp}+{\sf px},~$ and
$\big(-it/2+{\sf p}\big)^2+2\big(-it/2+{\sf p}\big)\big(it/2+{\sf p}\big)+\big(it/2+{\sf p}\big)^2=4{\sf p}^2.$

\medskip
\noindent
Together, Eqs. \eqref{eqn:Htt}, \eqref{eqn:Htk}, and \eqref{eqn:Hkk} yield 
\begin{align*}
\det(\mathcal{H}_{t,k}) &= \left( \ll {\sf x}^2 \gg -(\ll {\sf x} \gg)^2 \right) \left(\ll {\sf p}^2 \gg -(\ll {\sf p} \gg)^2 \right)
-\left({1 \over 2}\ll{\sf xp}+{\sf px} \gg  -\ll {\sf x} \gg \ll {\sf p} \gg \right)^2 \\
&=\langle \! \langle \big({\sf x}^2 -\ll {\sf x} \gg \!\!\big)^2 \rangle \! \rangle    \langle \! \langle \big({\sf p}^2 -\ll {\sf p} \gg \!\!\big)^2 \rangle \! \rangle -\left({\langle \! \langle \big({\sf x}^2 -\ll {\sf x} \gg \!\!\big)\big({\sf p}^2 -\ll {\sf p} \gg \!\big) \rangle \! \rangle+\langle \! \langle \big({\sf p}^2 -\ll {\sf p} \gg \!\! \big)\big({\sf x}^2 -\ll {\sf x} \gg \!\! \big) \rangle \! \rangle \over 2}\right)^2 \\
&\geq 0,
\end{align*}
by the Cauchy-Bunyakovsky-Schwarz inequality.

\end{proof}

\section{Rate Function for Husimi Distributions}
\label{app_Husimi_LDP}
In this section we present a complementary result covering the large deviation theory to Husimi distributions. Denote by $\varphi_\omega(x,p)$ the normal $\mathcal{N}(0,\Omega)$ density function with the covariance matrix given by
$$\Omega={1 \over 2}\left(\begin{array}{cc}{1 \over m \omega} & 0 \\0 & m\omega\end{array}\right).$$
The {\it Husimi distribution} is define as the following convolution
\be\label{eqn:Husimi}
H(x,p)=\varphi_\omega \ast W(x,p).
\ee
Therefore, its moment generating function
\be\label{eqn:HusimiMGF}
M_H(t,k)= \iint\limits_{\mathbb{R}^2} \! e^{tx+kp} \, H(x,p) \,dx\, dp=
\exp\left\{-{1 \over 4m \omega}t^2-{m \omega \over 4}k^2\right\} \, M_W(t,k),
\ee
and its log-moment generating function (i.e., the cumulant generating function) equals
\begin{align}\label{eqn:HusimiCGF}
\log M_H(t,k)&=\log M_W(t,k)+{1 \over 4m \omega}t^2+{m \omega \over 4}k^2 \nonumber \\
&=\log\langle e^{k{\sf p}/2} e^{t{\sf x}} e^{k{\sf p}/2} \rangle +{1 \over 4m \omega}t^2+{m \omega \over 4}k^2.
\end{align}
Thus, the {\it large deviations entropy rate} of $H(x,p)$, defined via Legendre transform,
\begin{align}\label{eqn:entropyHldp}
I_H(x,p)&=\sup\limits_{t,k \in \mathbb{R}}\big(xt+pk -\log M_H(t,k)\big) \nonumber \\
&=\sup\limits_{t,k \in \mathbb{R}}\left(xt+pk -{1 \over 4m \omega}t^2-{m \omega \over 4}k^2-\log \langle e^{k{\sf p}/2} e^{t{\sf x}} e^{k{\sf p}/2} \rangle\right) 
\nonumber \\
&\leq I_W(x,p).
\end{align}
From Eq. \eqref{eqn:SLDP} one can then define and calculate the large deviation principle entropy based on the Husimi distribution.
\section{Momentum Generating Function of the Harmonic Oscillator }
\label{app_generating_funciton_HO}
In this section we will present in more details the results presented in Section \ref{sec:Harmonic_Oscillator_Analytical_Numerical_Results}, for the momentum generating function of the 1D Harmonic Oscillator. Considering $|n \rangle$ as the Fock state representation for the harmonic oscillator, the generating function is then, 
\begin{equation}
    M^{(n)}_W (t,k) =  \langle n | e^{\frac{k p}{2}} e^{tx} e^{\frac{k p}{2}}   | n \rangle.
    \label{eq:exp1}
 \end{equation}
We make use of the momentum and position operator in term of the ladder operator,
\begin{equation}
    \begin{aligned}
        x = \sqrt{\frac{\hbar}{ 2 m \omega}} (a + a^{\dagger}) \qquad \text{and} \qquad p =-i \sqrt{\frac{m \omega \hbar}{2}} (a - a^{\dagger}).
    \end{aligned}
    \label{eq:ladderxp}
\end{equation}
By applying the Baker-Campbell-Hausdorff (BCH) identity $e^{A} e^{B} = e^{A + B} e^{\frac{1}{2} [A,B]}$ on Eq. \eqref{eq:exp1}, 
\begin{equation}
    \begin{aligned}
        M^{(n)}_W (t,k) &= e^{\frac{1}{2}  |\alpha|^{2}} \langle n | e^{\alpha  a^{\dagger}} e^{ \alpha ^{*} a}| n \rangle,
    \end{aligned}
    \label{eq:step2}
\end{equation}
where we have defined $\alpha =\bar{t} + i \bar{k} $, $\alpha^{*} = \bar{t} - i \bar{k}$, $\bar{t} = t \sqrt{\frac{\hbar}{ 2 m \omega}}$ with $\bar{k} = k \sqrt{\frac{m \omega \hbar}{2}}  $ yielding,
\begin{equation}
   \begin{aligned}
        |\alpha|^{2} &=  (\bar{t} - i \bar{k})(\bar{t} + i \bar{k}) = t^{2} \frac{\hbar}{2 m \omega} + k^{2}\frac{m \omega \hbar}{2}.
   \end{aligned}
   \label{eq:alpha2def}
\end{equation}
To study the action of the ladder operators in the Fock state, we use that, 
\begin{equation}
        e^{\alpha a^{\dagger}} = \sum_{j = 0}^{\infty} \frac{\alpha^{j}}{j!} (a^{\dagger})^{j}\qquad \text{and,} \qquad  e^{\alpha^{*} a} = \sum_{q = 0}^{\infty} \frac{(\alpha^{*})^{q}}{q!} a^{q} ,
\end{equation}
inserting into Eq.~(\ref{eq:step2}), 
\begin{equation}
    \begin{aligned}
        M^{(n)}_W (t,k) = e^{\frac{1}{2}  |\alpha|^{2}} \sum_{j = 0}^{\infty} \sum_{q = 0}^{\infty} \frac{(\alpha^{*})^{q} \alpha^{j}}{q! j!} \langle n |(a^{\dagger})^{j}  a^{q} | n \rangle. \\
    \end{aligned}
    \label{eq:Mladder}
\end{equation}
The ladder operator on $|n\rangle$, 
\begin{equation}
    \begin{aligned}
        a^{q} |n\rangle &=  \sqrt{\frac{n!}{(n-q)!}} |n - q\rangle \qquad &\text{for} \qquad q \leq n \\
        \langle n | a^{\dagger j} & =  (a^{j} | n \rangle)^{\dagger} = \left( \sqrt{\frac{n!}{(n-j)!}} |n - j\rangle  \right)^{\dagger}  =  \sqrt{\frac{n!}{(n-j)!}} \langle n - j | \qquad &\text{for} \qquad j \leq n.
    \end{aligned}
    \label{eq:ladderfock}
\end{equation}
Inserting Eq. \eqref{eq:ladderfock} into Eq. \eqref{eq:Mladder} yields, 
\begin{equation}
     \begin{aligned}
         M_W^{(n)} = e^{\frac{1}{2}  |\alpha|^{2}} \sum_{q = 0}^{n} \frac{|\alpha|^{2 q}}{(q!)^{2}}\frac{n!}{(n-q)! q!} = e^{\frac{1}{2}  |\alpha|^{2}} \sum_{q = 0}^{n} \binom{n}{q} \frac{|\alpha|^{2 q}}{q!},
     \end{aligned}
\end{equation}
which has the similar representation of the Laguerre polynomials, and therefore
\begin{equation}
     M_W^{(n)}(t,k) = e^{\frac{1}{2}  |\alpha|^{2}} \text{L}_n (- |\alpha|^{2}).
\end{equation}
Substituting the $|\alpha|^{2}$ in Eq. \eqref{eq:alpha2def} into the above expression,
\begin{equation}
     M_W^{(n)}(t,k) = e^{t^{2} \frac{\hbar}{4 m \omega} + k^{2}\frac{m \omega \hbar}{4}} \text{L}_n\left(- t^{2} \frac{\hbar}{2 m \omega} - k^{2}\frac{m \omega \hbar}{2} \right).
     \label{eq:generalMn_HO}
\end{equation}
This is the general form of the momentum generating function for the n-th state of the harmonic oscillator.
\section{Proof of the Preposition I}
\label{app_proof_prep_witness}
\begin{proof}
If $\rho_{AB}=\rho_A\otimes\rho_B$, then the Wigner distribution and its generating function factorize:
\begin{align}
W_{AB}(r_A,r_B)&=W_A(r_A)W_B(r_B),\\
M_{AB}(s_A,s_B)&=M_A(s_A)M_B(s_B),
\end{align}
consequently, the cumulant generating function
\begin{equation}
\log M_{AB}(s_A,s_B) = \log M_A(s_A)+ \log M_B(s_B),
\end{equation}
and the rate functions,
\begin{equation}
I_{AB}(r_A,r_B)=I_A(r_A)+I_B(r_B).
\end{equation}
It follows that $P_{AB,\mathrm{LDP}}=P_{A,\mathrm{LDP}}P_{B,\mathrm{LDP}}$ and, by definition of the mutual information Eq. \eqref{eqn:MI-LDP-def}, $\mathcal I_{\mathrm{LDP}}^{(W)}(A:B)=0$. By the uniqueness of the generating function,
\begin{equation}
W_{\rho_A\otimes\rho_B}(r_A,r_B) =W_{\rho_A}(r_A)W_{\rho_B}(r_B).
\end{equation}
consequently,
\begin{equation}
W_{\rho_{AB}}=W_{\rho_A\otimes\rho_B} \quad\Longrightarrow\quad \rho_{AB}=\rho_A\otimes\rho_B.
\end{equation}
Finally, for a globally pure bipartite state, being separable is
equivalent to being a product state. Conversely,
\begin{equation}
\rho_{AB}\neq\rho_A\otimes\rho_B \quad\Longleftrightarrow\quad \rho_{AB}\ \text{is entangled},
\end{equation}
in which produce $\mathcal I_{\mathrm{LDP}}^{(W)}(A:B)>0$. 
\end{proof}

\end{document}